\documentclass[aps,prb,preprint,groupedaddress,showpacs,amsmath,amssymb]{revtex4}

\usepackage{graphicx}
\usepackage{dcolumn}
\usepackage{bm}

\begin{document}

\title{Heat capacity jump at $T_c$  and pressure derivatives of superconducting transition temperature in the Ba$_{1-x}$Na$_x$Fe$_2$As$_2$ ($0.1 \leq x \leq 0.9$) series}

\author{Sergey L. Bud'ko$^{1}$, Duck Young Chung$^{2}$, Daniel Bugaris$^{2}$, Helmut Claus$^2$, Mercouri G. Kanatzidis$^{2,3}$, and Paul C. Canfield$^{1}$}
\affiliation{$^{1}$Ames Laboratory, US DOE and Department of Physics and Astronomy, Iowa State University, Ames, Iowa 50011, USA}
\affiliation{$^{2}$Materials Science Division, Argonne National Laboratory, Argonne, Illinois 60439-4845, USA}
\affiliation{$^{3}$Department of Chemistry, Northwestern University, Evanston, Illinois 60208-3113, USA}

\date{\today}

\begin{abstract}

We present the evolution of the initial (up to $\sim 10$ kbar) hydrostatic, pressure dependencies of $T_c$ and of the ambient pressure jump in the heat capacity associated with the superconducting transition as a function of Na - doping in the  Ba$_{1-x}$Na$_x$Fe$_2$As$_2$ family of iron-based superconductors. For Na concentrations $0.15 \leq x \leq 0.9$, the jump in specific heat at $T_c$,  $\Delta C_p|_{T_c}$, follows the $\Delta C_p \propto T^3$ scaling found for most BaFe$_2$As$_2$ - based superconductors. Pressure dependencies are non-monotonic for $x = 0.2$ and $x = 0.24$. For other Na concentrations $T_c$ decreases under pressure in almost linear fashion. The anomalous behavior of the $x = 0.2$ and $x = 0.24$ samples under pressure are possibly due to the crossing of the phase boundaries of the narrow antiferromagnetic tetragonal phase, unique for the Ba$_{1-x}$Na$_x$Fe$_2$As$_2$ series, with the application of pressure.

\end{abstract}

\pacs{74.62.Fj, 74.70.Xa, 74.25.Bt, 74.62.Dh}

\maketitle

\section{Introduction}

Of many recently discovered Fe-based superconductiors and related materials \cite{joh10a,ste11a,joh11a} the so called 122 family ($AE$Fe$_2$As$_2$ with $AE = $ alkaline earth and Eu), is the most studied. \cite{can10a,nin11a,man10a} This family allows for substitution on all three cystallographic sites and, as a result, an intricate combination of carrier-doping and anisotropic steric effects can be studied, while maintaining the same, tetragonal, ThCr$_2$Si$_2$ - type crystal structure. 

The main body of the published work on the 122 family has been focused on the $AE$(Fe$_{1-x}$$TM_x$)$_2$As$_2$ series with transition metals, $TM$, being substituted for Fe.\cite{can10a,nin11a}  This is due to the relative ease of growing homogeneous, high quality single crystals. Substitutions for $AE$, as in the Ba$_{1-x}$K$_x$Fe$_2$As$_2$ series, \cite{rot08a,rot08b,che09a,avc12a}  or for As as in the BaFe$_2$(As$_{1-x}$P$_x$)$_2$ series \cite{jia09a,kas10a} have been explored, but both series require significant efforts to achieve homogeneity and/or reasonable size of the crystals.

At first glance, substitution for Ba [Ba$_{1-x}$K$_x$Fe$_2$As$_2$], Fe [Ba(Fe$_{1-x}$Co$_x$)$_2$As$_2$] or As [BaFe$_2$(As$_{1-x}$P$_x$)$_2$] in BaFe$_2$As$_2$, as well as application of pressure, \cite{col09a} to this parent compound result in similar phase diagrams. \cite{pag10a} First, the temperature of structural and magnetic transitions decreases, then superconductivity emerges with a a region of coexistence of superconductivity and antiferromagnetism. On further substitution (or under higher pressure) the  magnetic and structural transitions are suppressed, the superconducting transition temperature passes through the maximum  and gradually goes to zero, or to a small finite value as in the case of complete substitution of K for Ba, in KFe$_2$As$_2$. Closer examination of the globally similar phase diagrams though, point to clear differences in details that allow to gain an insight into the complex physics of these materials. \cite{ste11a,can10a,nin11a,man10a}

Na - substitution for the $AE$ appears to be one of the less - explored branches in the 122 family tree. A possibility to induce superconductivity by Na-substitution in CaFe$_2$As$_2$ was realized fairly early,  \cite{shi08a,wug08a} whereas a tentative $x - T$ phase diagram for the Ca$_{1-x}$Na$_x$Fe$_2$As$_2$ series was published a few years later. \cite{zha11a} A Sr$_{1-x}$Na$_x$Fe$_2$As$_2$ sample with $T_c$ value as high as $\approx 35$ K was studied in Ref. \onlinecite{gok09a}, and later an evolution of the physical properties in Sr$_{1-x}$Na$_x$Fe$_2$As$_2$ (for a rather limited range of substitution, $x \leq 0.4$) was presented. \cite{cor11a}

The Ba$_{1-x}$Na$_x$Fe$_2$As$_2$ series, where Na is substituted for Ba, offers an almost complete range of substitution \cite{cor10a,avc13a}. One of the complications for this series is that its end member, NaFe$_2$As$_2$ ($T_c \sim 11 - 12$ K), was reported to be metastable, and as such cannot be formed by a solid-state reaction technique, but can only be obtained by the mild oxidation of NaFeAs. \cite{tod10a,goo10a,fri12a} Additionally, deviations from stoichiometry (Na$_{1-y}$Fe$_{2-x}$As$_2$, with $y \approx 0.1$ and $x \approx 0.3$) for the obtained material were suggested.\cite{fri12a} 

In generic terms, for the overlapping $x$ - values, the $x - T$ phase diagram of the  Ba$_{1-x}$Na$_x$Fe$_2$As$_2$ series \cite{avc13a} bears a close similarity to that of the Ba$_{1-x}$K$_x$Fe$_2$As$_2$ series, \cite{rot08b,avc12a} but a comprehensive study of the Na - doped series and a direct comparison with the results for its  K - doped counterpart would help to address several issues of relevance for the physics of Fe-based superconductors. First, in the  Ba$_{1-x}$K$_x$Fe$_2$As$_2$ series several experimental observations point to a significant modification of the superconducting state (possibly change in superconducting pairing symmetry) for K - concentration, $x > 0.7$. \cite{hir12a,mal12a,bud13a,wat13a,ota13a} It would be of interest to examine the overdoped part of the phase diagram of the  Ba$_{1-x}$Na$_x$Fe$_2$As$_2$ series for similar features. Second, in addition to the low temperature antiferromagnetic / orthorhombic phase, that is ubiquitous  in Fe-based superconductors, an antiferromagnetic tetragonal, $C4$, phase was reported in the  Ba$_{1-x}$Na$_x$Fe$_2$As$_2$ series over a narrow Na - concentration region around $x \sim 0.24$. \cite{avc13a,avc13b} The tip of the narrow $C4$ dome was suggested to be at $\sim 50$ K; at lower temperatures this new magnetic phase was suggested to co-exist with superconductivity. The bulk physical properties of the members of the series close to and in the $C4$ phase as well as the effect of this magnetic phase on superconductivity are interesting to study in detail.

In this work we utilize the same approach as we did in the recent studies of the Ba$_{1-x}$K$_x$Fe$_2$As$_2$ series; \cite{bud13a}  we present two set of data for the Ba$_{1-x}$Na$_x$Fe$_2$As$_2$ series, with Na - concentrations covering underdoped, optimally doped and overdoped regions of the  $x - T$ phase diagram.   The first set consists of the data on the evolution of the jump in heat capacity at the superconducting transition. Many  Fe - based, 122 superconductors follow the empirical trend suggested in Ref. \onlinecite{bud09b} and expanded in Refs. \onlinecite{kim11b,kim12a}, the so-called BNC scaling, $\Delta C_p|_{T_c} \propto T_c^3$. This set is to be compared with the data on the  Ba$_{1-x}$K$_x$Fe$_2$As$_2$ series, \cite{bud13a} that show clear deviation from the BNC scaling for K - concentration $x > 0.7$. The second set contains the initial ($P \lesssim 10$ kbar) pressure dependencies of the superconducting transition temperatures, $T_c(P)$ similar in scope to the data reported for Ba(Fe$_{1-x}$Co$_x$)$_2$As$_2$ \cite{ahi09a} and Ba$_{1-x}$K$_x$Fe$_2$As$_2$. \cite{bud13a} Such data evaluate possible equivalence of pressure and doping that was suggested for several 122 series. \cite{kim09a,dro10a,kli10a,kim11a} Moreover, under favorable circumstances such a dataset can shed light on the details of the mechanism of superconductivity. \cite{xio92a,cao95a,fie96a} 

\section{Experimental details}

Homogeneous, single phase Ba$_{1-x}$Na$_x$Fe$_2$As$_2$ polycrystalline powder with nominal $x = 0.1, 0.15, 0.2, 0.24, 0.3, 0.35, 0.4, 0.5, 0.6, 0.7, 0.8$, and 0.9 were synthesized by a variation of a previously reported procedure. \cite{avc13a}  Handling of all starting materials was performed in an Ar-filled glove box.  Mixtures of Ba, Na, and FeAs were loaded in alumina crucibles, which were sealed in Nb tubes under argon, and then sealed again in fused silica tubes under vacuum.  The mixtures were heated twice at  $800\,^{\circ}\mathrm{C}$ for 1 and 3 days respectively, and at  $850\,^{\circ}\mathrm{C}$ for 1 day, with quenching in air.  Between each annealing cycle, the mixtures were ground to a fine, homogeneous powder.  After the third annealing step, the powder was pressed into a pellet,   re-loaded into the alumina crucible, sealed first inside the Nb tube, and then into a fused silica tube.  The pellet was heated at  $850\,^{\circ}\mathrm{C}$ for approximately 1 day, before quenching in air.  The superconducting transition temperatures of the samples, as determined by magnetization measurements, were compared with those in the previously reported $x - T$ phase diagram \cite{avc13a} in order to confirm the $x$ values of the compositions.

The synthesis of homogeneous samples of Ba$_{1-x}$Na$_x$Fe$_2$As$_2$ is challenging due to volatility of Na upon heating.  For this reason, sealed Nb tubes are utilized to suppress and contain Na vaporization at high temperature.  This is particularly crucial in the highly underdoped region ($x < 0.3$) where small changes in $x$ can lead to large changes in $T_c$.  Because of the sensitivity of $T_c$ in the highly underdoped samples with respect to Na content, the pellets of these samples needed to be annealed multiple times for shorter durations in order to achieve sharp superconducting transitions.  Additionally, for the overdoped samples, a small amount of Na was added to the pelletized mixture before the final annealing in order to finely tune the $x$ parameter.  The most highly overdoped sample, Ba$_{0.1}$Na$_{0.9}$Fe$_2$As$_2$, is also difficult to prepare due to its tendency to phase-segregate into an underdoped composition with higher thermal stability and a stable ternary compound NaFeAs (with a different structure type). \cite{par09a} It should be noted here that the end-member NaFe$_2$As$_2$ ($x = 1$) cannot be formed by this solid-state technique, instead it can be obtained by the mild oxidation of NaFeAs. \cite{tod10a,goo10a,fri12a} 

Low-field dc magnetization under pressure, was measured in a Quantum Design Magnetic Property Measurement System, MPMS-5, SQUID magnetometer using a  a commercial, HMD, Be-Cu piston-cylinder pressure cell. \cite{hmd}  Daphne oil 7373 was used as a pressure medium and superconducting Pb or Sn (to have its superconducting transition well separated from that of the sample) as a low-temperature pressure gauge. \cite{eil81a}.  The heat capacity was measured using a hybrid adiabatic relaxation technique of the heat capacity option in a Quantum Design Physical Property Measurement System,  PPMS-14, instrument.

\section{Results}

\subsection{Jump in specific heat and BNC scaling}

\begin{figure}
\begin{center}
\includegraphics[angle=0,width=120mm]{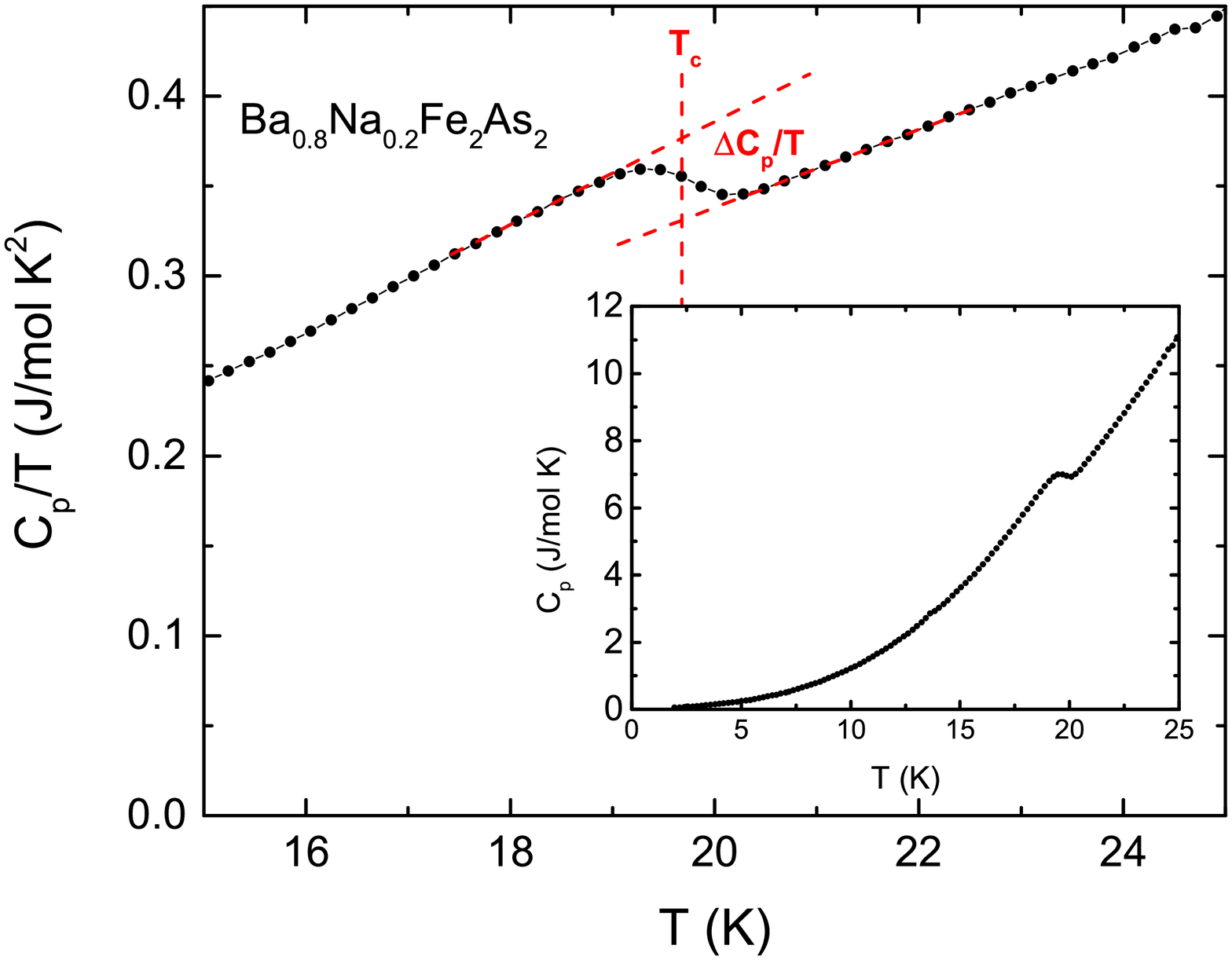}
\end{center}
\caption{Temperature-dependent heat capacity of  Ba$_{0.2}$Na$_{0.8}$Fe$_2$As$_2$ near the superconducting transition plotted as $C_p/T$ vs $T$. Criteria for $T_c$ and  $\Delta C_p|_{T_c}$ (isoentropic construct) are shown. Inset: $C_p(T)$ in a wider temperature range. } \label{F1}
\end{figure}

\begin{figure}
\begin{center}
\includegraphics[angle=0,width=120mm]{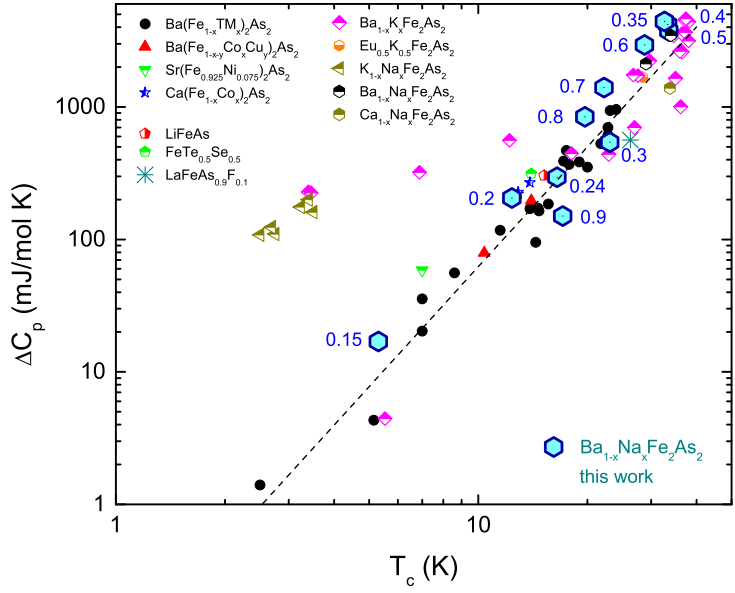}
\end{center}
\caption{(Color online) $\Delta C_p$ at the superconducting transition vs $T_c$  for the  Ba$_{1-x}$Na$_x$Fe$_2$As$_2$ series, plotted together with literature data for various FeAs-based superconducting materials. Plot from [\onlinecite{bud13a}] was updated to include  published data for K$_{1-x}$Na$_x$Fe$_2$As$_2$ ($0 \leq x \leq 0.31$), Ca$_{1-x}$Na$_x$Fe$_2$As$_2$, Ba$_{1-x}$Na$_x$Fe$_2$As$_2$ ($x = 0.35, 0.4$), and LaFeAs$_{0.9}$F$_{0.1}$. \cite{zha11a,pra11a,asw12a,abd13a,gri13a}  The line corresponds to $\Delta C_p \propto T_c^3$. Numbers near the symbols are Na - concentrations $x$.} \label{F2}
\end{figure}

Of the samples studied in this work, Ba$_{0.9}$Na$_{0.1}$Fe$_2$As$_2$ is not superconducting, whereas  Ba$_{0.15}$Na$_{0.85}$Fe$_2$As$_2$ is superconducting, but low field, zero field cooled, magnetization measurements display two transitions. The $x = 0.1$ sample is not included in the analysis of the jump in specific heat at $T_c$. For $x = 0.15$ the feature at $\sim 5$ K is the only feature present in the specific heat data. For this sample the difference between the data taken in zero field and 140 kOe applied field was analyzed and the error bars in the obtained $\Delta C_p$ at $T_c$ value are expected to be rather large. The other samples studied in this work  show a distinct feature in specific heat at $T_c$ (see Fig. \ref{F1} as an example). The $T_c$  and  $\Delta C_p|_{T_c}$ values were determined by a procedure consistent with that used in Ref. \onlinecite{bud13a,bud09b}. The specific heat jump data for  the  Ba$_{1-x}$Na$_x$Fe$_2$As$_2$ series obtained in this work were added in Fig. \ref{F2} to the BNC plot taken from Ref. \onlinecite{bud13a} and updated by including several more recent data points taken from the evolving literature. There appears to be a clear trend: all data points ($0.15 \leq x \leq 0.9$) from this study follow the BNC scaling, in agreement with the literature data for two close Na-concentrations. \cite{pra11a,asw12a} This behavior is clearly different from that reported in the Ba$_{1-x}$K$_x$Fe$_2$As$_2$ series, where the data for K - concentrations of $0.8 \leq x \leq 1.0$ clearly deviate from the BNC scaling, and is similar to previously the studied Ba(Fe$_{1-x}${\it TM}$_x$)$_2$As$_2$ ({\it TM} = transition metal) series, for which the BNC scaling was observed for the samples covering the full extent of the superconducting dome. For the Ba$_{1-x}$Na$_x$Fe$_2$As$_2$ series, $\Delta C_p$ at $T_c$ increases and decreases as $T_c$ rises and falls to form the superconducting dome.

\subsection{Pressure dependence of $T_c$}

\begin{figure}
\begin{center}
\includegraphics[angle=0,width=120mm]{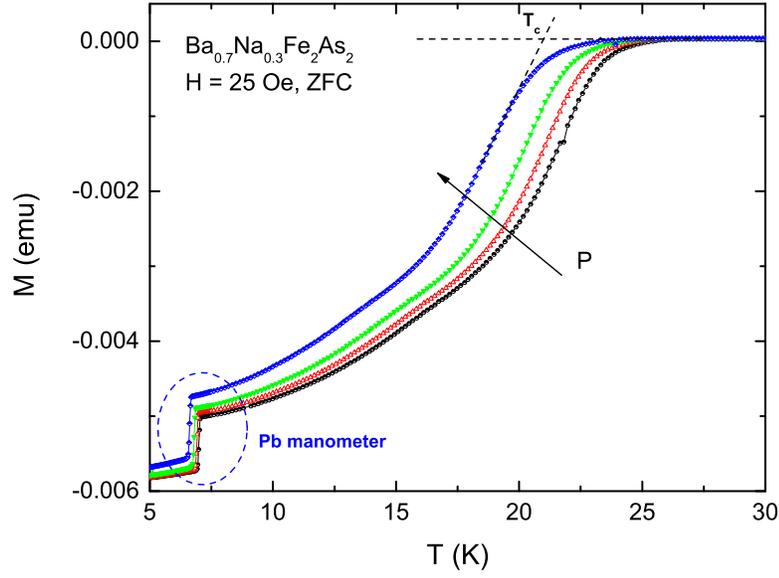}
\end{center}
\caption{(Color online)   Example of temperature dependent magnetization (zero-field-cooled, taken in 25 Oe applied magnetic field) of Ba$_{0.7}$Na$_{0.3}$Fe$_2$As$_2$ measured at 1.1, 2.9, 5.9, and 10.7 kbar. Arrow indicates increasing pressure. The onset criterion for $T_c$ used in this work is shown for $P = 10.7$ kbar curve as an example. Superconducting transitions  in Pb used as a pressure gauge are seen near 7 K.} \label{F3}
\end{figure}

\begin{figure}
\begin{center}
\includegraphics[angle=0,width=120mm]{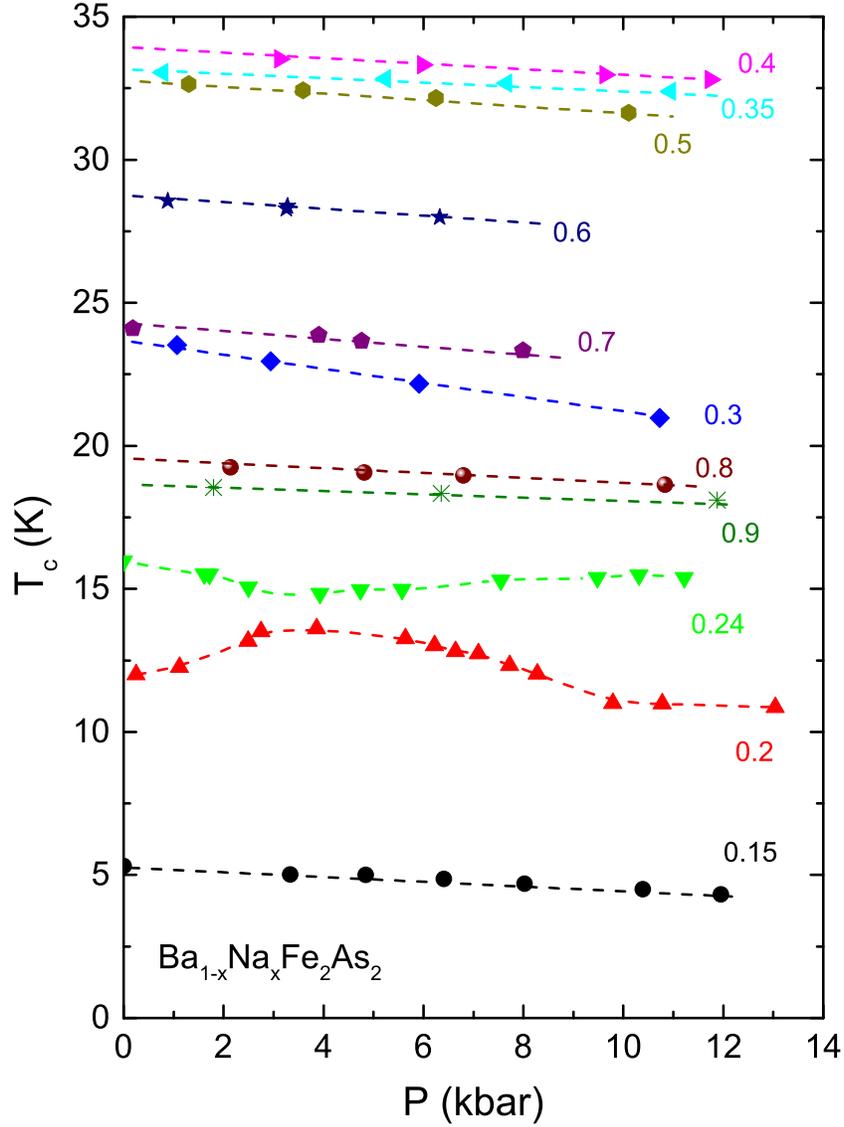}
\end{center}
\caption{(Color online) Summary plot of the pressure dependence of $T_c$ for the Ba$_{1-x}$Na$_x$Fe$_2$As$_2$ ($0.15 \leq x \leq 0.9$) series studied in this work. Dashed lines are linear fits to the data except for the $x = 0.2$ and $x = 0.24$ where they are guide for the eye. The linear fits are extended to $P = 0$.} \label{F4}
\end{figure}

An example of $M(T)$ data taken at different pressures is shown in Fig. \ref{F3}. An onset criterion was used to determine $T_c$. Alternative criteria (e.g. maximum in $dM/dT$) yield similar pressure dependencies.  The  Ba$_{0.9}$Na$_{0.1}$Fe$_2$As$_2$ sample was measured at ambient pressure and at 11.1 kbar. No traces of superconductivity down to 1.8 K were observed. The pressure dependencies of the superconducting transition temperatures of the samples with Na - concentration in the range of $0.15 \leq x \leq 0.9$ are shown in Fig. \ref{F4}. For the Ba$_{1-x}$Na$_x$Fe$_2$As$_2$ sample with Na concentration of $x = 0.15$, the pressure dependence associated with the lower temperature feature, that is consistent with the ambient pressure phase diagram, is presented. \cite{avc13a} It is noteworthy that (i) $T_c(P)$ are non-monotonic (even in a limited pressure range of this work) for two Na concentrations, $x = 0.2$ and $x = 0.24$; (ii) for all other concentrations studied in this work, both in underdoped and overdoped regimes, $T_c$ decreases under pressure. The data for Ba$_{1-x}$Na$_x$Fe$_2$As$_2$ samples with $x = 0.2$ and $x = 0.24$ are shown separately in Fig. \ref{F5} (a) and (b). Two pressure runs on two different samples from the same batch were performed for  Ba$_{0.8}$Na$_{0.2}$Fe$_2$As$_2$. The results are faily consistent. For both, $x = 0.2$ and $x = 0.24$, samples the measurements were performed on pressure increase and pressure decrease ( Fig. \ref{F5} (a), (b)). For both samples the observed non-monotonic behavior is robust and not affected by pressure cycling. One can join these two data sets  by shifting the data for Ba$_{0.76}$Na$_{0.24}$Fe$_2$As$_2$  by $+ 8$ kbar along the $X$ - axis and by $- 4$ K along the $Y$ - axis. We can understand the grounds for such two-axis shift if we assume that both, steric effect and hole doping, cause changes in superconducting transition temperature.

\begin{figure}
\begin{center}
\includegraphics[angle=0,width=120mm]{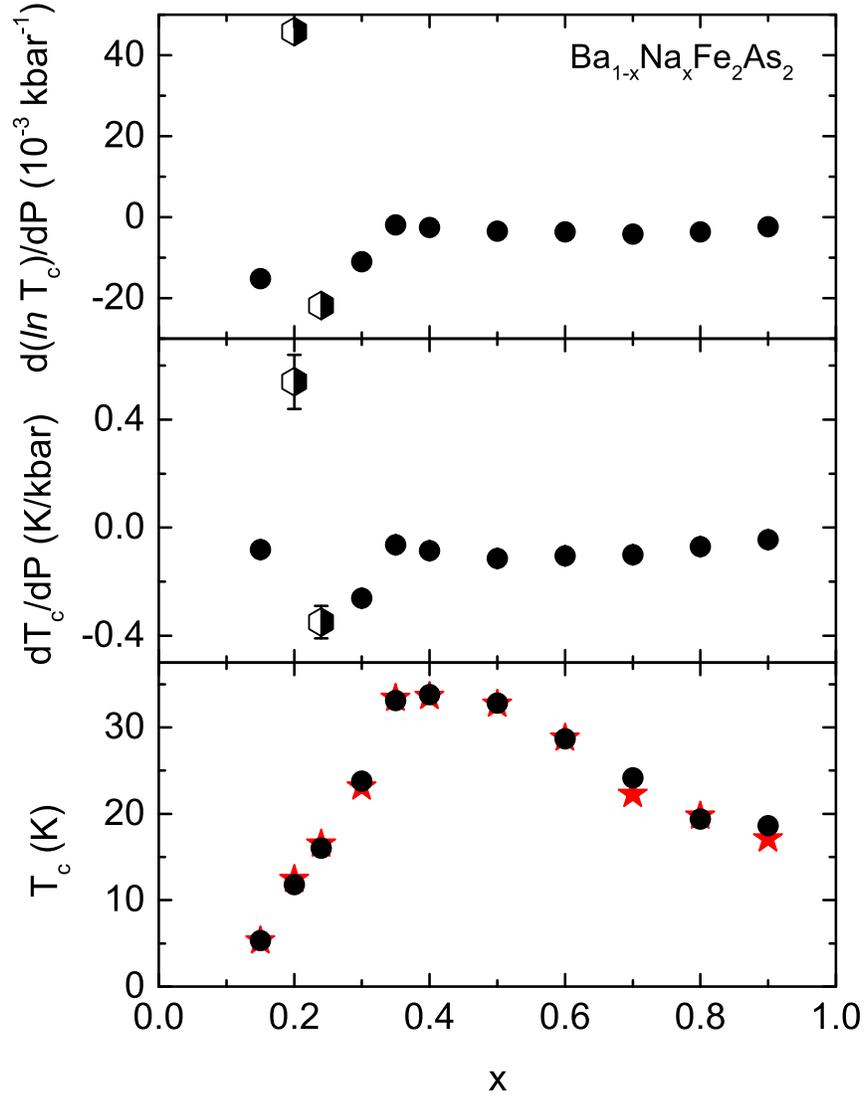}
\end{center}
\caption{(Color online) $d(\ln T_c)/dP$, $d T_c/dP$ and $T_{c0}$ (top to bottom) as a function of Na-concentration in Ba$_{1-x}$Na$_x$Fe$_2$As$_2$. Stars: $T_c$ values from heat capacity measurements. For  Ba$_{0.8}$Na$_{0.2}$Fe$_2$As$_2$ and Ba$_{0.76}$Na$_{0.24}$Fe$_2$As$_2$ initial, low pressure, pressure derivatives' values are used (half-filled hexagons).} \label{F6}
\end{figure}

A more compact way to look at the pressure dependence of $T_c$ in the  Ba$_{1-x}$Na$_x$Fe$_2$As$_2$ series is presented in Fig. \ref{F6}. The superconducting transition temperature values obtained from specific heat and magnetization measurements are very similar and these data sets are consistent with the literature. \cite{avc13a} The initial, low pressure, value of $d T_c/dP$ (and $d(\ln T_c)/dP$) for $x = 0.2$ is positive and relatively high. For other Na - concentrations studied the pressure derivatives of $T_c$ are negative. Whereas for optimally doped and overdoped the absolute values of $d T_c/dP$ and $d(\ln T_c)/dP$ are rather small and change smoothly with concentration (Fig. \ref{F6}), there appears to be a break of the trend in the underdoped region.  

\begin{figure}
\begin{center}
\includegraphics[angle=0,width=120mm]{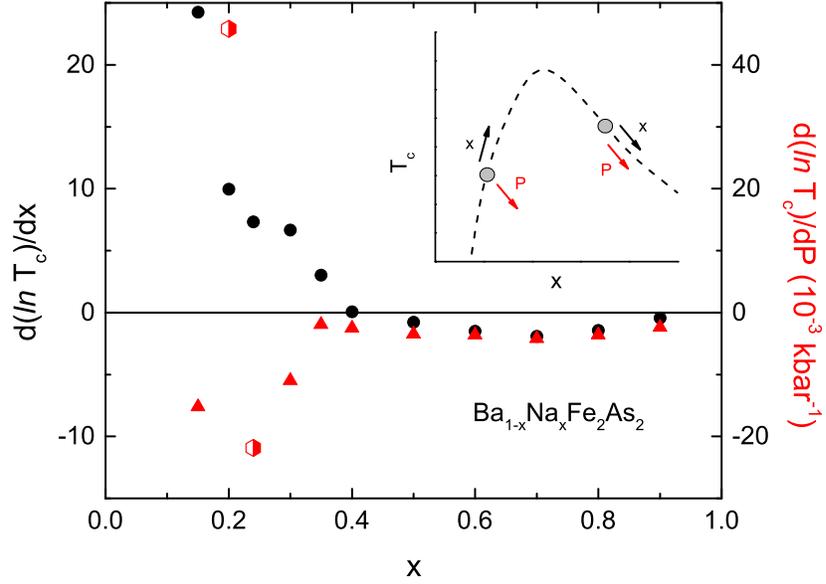}
\end{center}
\caption{(Color online) Na - concentration dependence of the normalized concentration derivatives, $d(\ln T_c)/dx = \frac{1}{T_{c0}}~d T_c/dx$ (left axis, circles), and the normalized pressure derivatives, $d(\ln T_c)/dP = \frac{1}{T_{c0}}~d T_c/dP$ (right axis, triangles)  of the superconducting transition temperatures. For  Ba$_{0.8}$Na$_{0.2}$Fe$_2$As$_2$ and Ba$_{0.76}$Na$_{0.24}$Fe$_2$As$_2$  initial, low pressure, normalized pressure derivatives' values are used (half-filled hexagons). Inset: schematic exhibiting {\it different} signs of the change in $T_c$ with substitution and with pressure for the underdoped samples, and {\it similar} effects of substitution and pressure for the overdoped samples.} \label{F7}
\end{figure}

Fig. \ref{F7} presents a comparison of the relative changes in superconducting transition temperature under pressure and with Na - doping.   For $0.4 \leq x \leq 0.9$ both sets of data can be scaled reasonably well, illustrating apparent equivalence of the effect of pressure and doping on $T_c$, suggested for other members of the 122 family \cite{kim09a,dro10a,kli10a,kim11a} and also observed in the limited range of K - concentrations for a closely related  Ba$_{1-x}$K$_x$Fe$_2$As$_2$. \cite{bud13a} This scaling however fails for Na concentrations $0.15 < x \leq 0.35$. Not just the values of  $d(\ln T_c)/dP$ and  $d(\ln T_c)/dx$ cannot be scaled in this region of concentrations, but (except for $x = 0.2$) the signs of these derivatives are different. In the underdoped region increase in $x$ causes an increase in $T_c$, and pressure causes decrease in $T_c$, however in the optimally doped and overdoped regions both, increase in $x$ and pressure cause decrease in $T_c$ (Fig. \ref{F7}, inset).

\section{Discussion and summary}

Both K and Na substitutions in BaFe$_2$As$_2$  provide hole doping and induce superconductivity with comparable maximum values of $T_c$ of $34 - 38$ K at similar K or Na concentrations of $x \approx 0.4$. 

Whereas in the  Ba$_{1-x}$K$_x$Fe$_2$As$_2$ series a clear deviation from the BNC scaling is observed for $0.8 \leq x \leq 1$, \cite{bud13a} in the  Ba$_{1-x}$Na$_x$Fe$_2$As$_2$ series the data for $0.15 \leq x \leq 0.9$ follow the BNC scaling,  $\Delta C_p|_{T_c} \propto T_c^3$, fairly well. This probably means that either there is no significant modification of the superconducting state (e.g. change in superconducting gap symmetry) in the Ba$_{1-x}$Na$_x$Fe$_2$As$_2$ series over the whole studied Na concentration range, or, if such modification exists, it is very subtle in its implications for the BNC scaling. The fact that the  Ba$_{1-x}$Na$_x$Fe$_2$As$_2$ series does not extend, in single phase form, to $x = 1.0$ prevents us from carrying this study to pure NaFe$_2$As$_2$, as we were able to for KFe$_2$As$_2$.

The negative sign of the pressure derivatives of $T_c$ for the underdoped samples (except for $x = 0.2$) is a clear indication of the non-equivalence of substitution and pressure for the Ba$_{1-x}$Na$_x$Fe$_2$As$_2$ series in this range, that is different from the gross overall equivalence suggested for other 122 series. \cite{bud13a,kim09a,dro10a,kli10a,kim11a} It has to be noted that for the  Ba(Fe$_{1-x}${\it TM}$_x$)$_2$As$_2$ ({\it TM} = transition metal)  (at least for {\it TM} concentrations that cover the superconducting dome) \cite{nin09a,rul10a,tha10a} and for Ba$_{1-x}$K$_x$Fe$_2$As$_2$ series \cite{rot08b,avc12a} the concentration dependence of the lattice parameters is monotonic and close to linear.  For the  Ba$_{1-x}$Na$_x$Fe$_2$As$_2$ series, the $a$ lattice parameter decreases with increase of $x$  in almost linear fashion, but the $c$ lattice parameter initially increases and then decreases, with  a maximum  at $x \sim 0.4$ in its dependence of Na - concentration. \cite{avc13a}. Although we do not know which particular structural parameter in the Ba$_{1-x}$Na$_x$Fe$_2$As$_2$ series has the dominant contribution to the pressure dependence of $T_c$, this non-monotonic behavior of $c(x)$ might be responsible for the negative sign of the $dT_c/dP$ for underdoped samples. Detailed structural studies under pressure would be useful for deeper understanding of this problem.

\begin{figure}
\begin{center}
\includegraphics[angle=0,width=82.5mm]{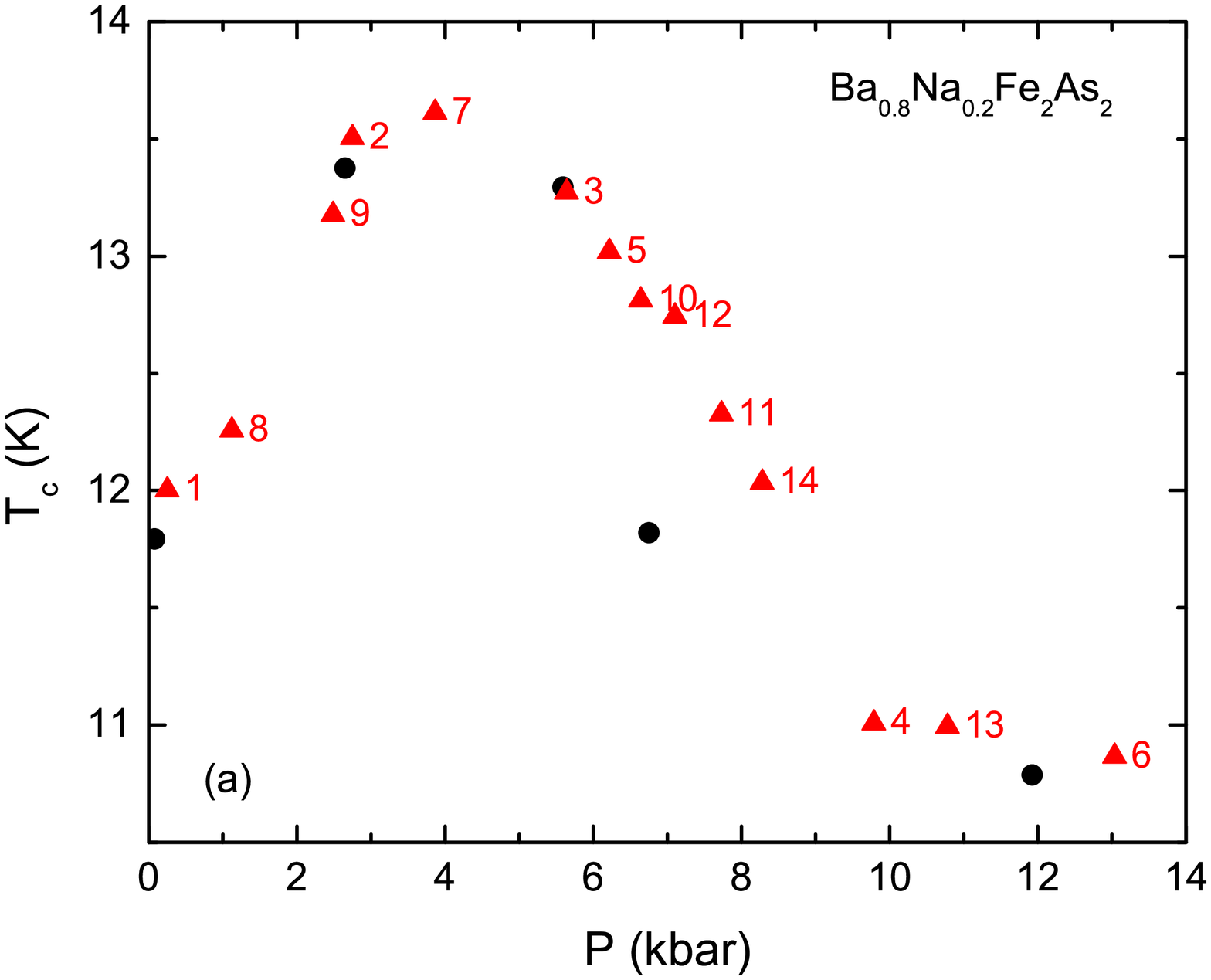}
\includegraphics[angle=0,width=82.5mm]{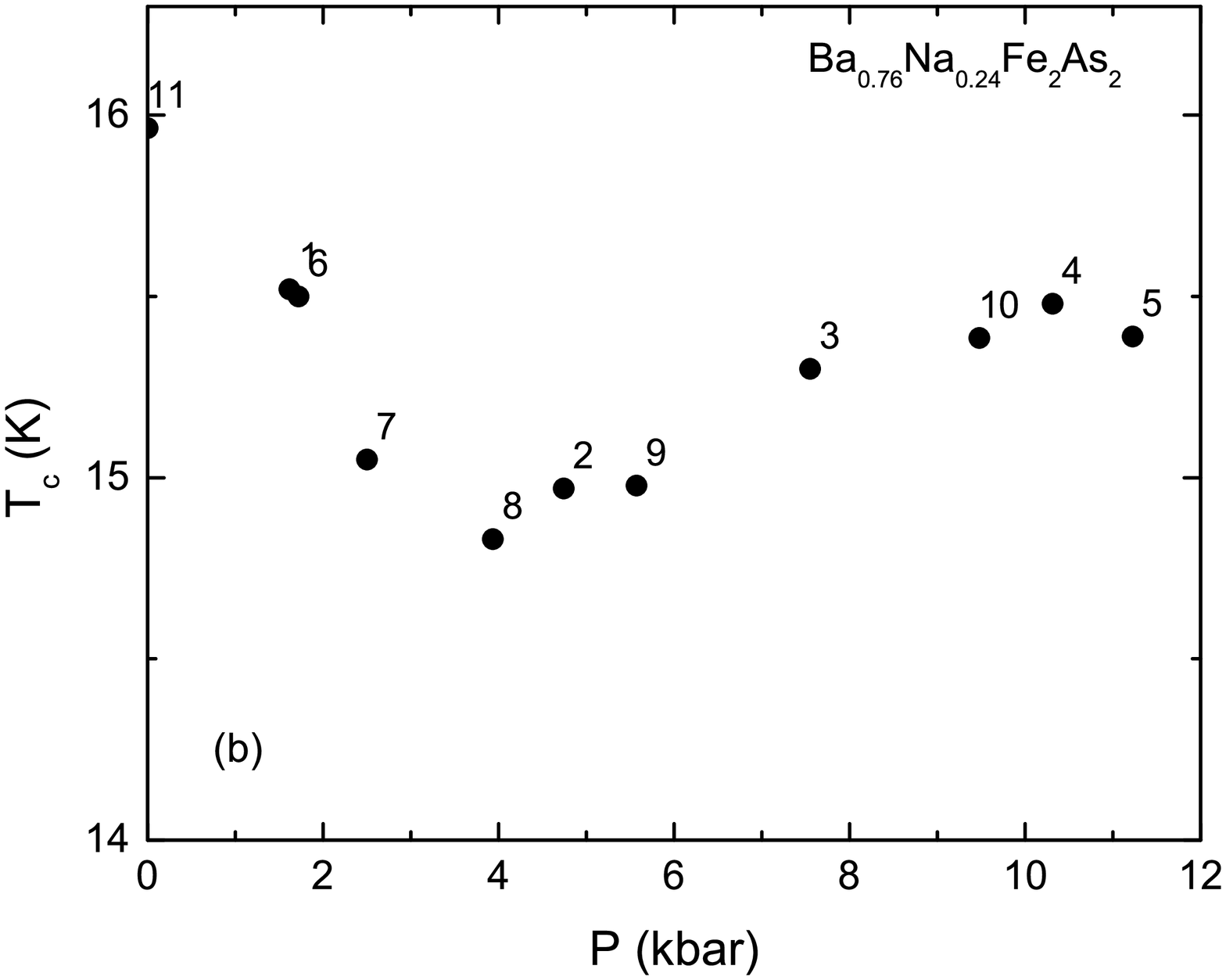}
\includegraphics[angle=0,width=82.5mm]{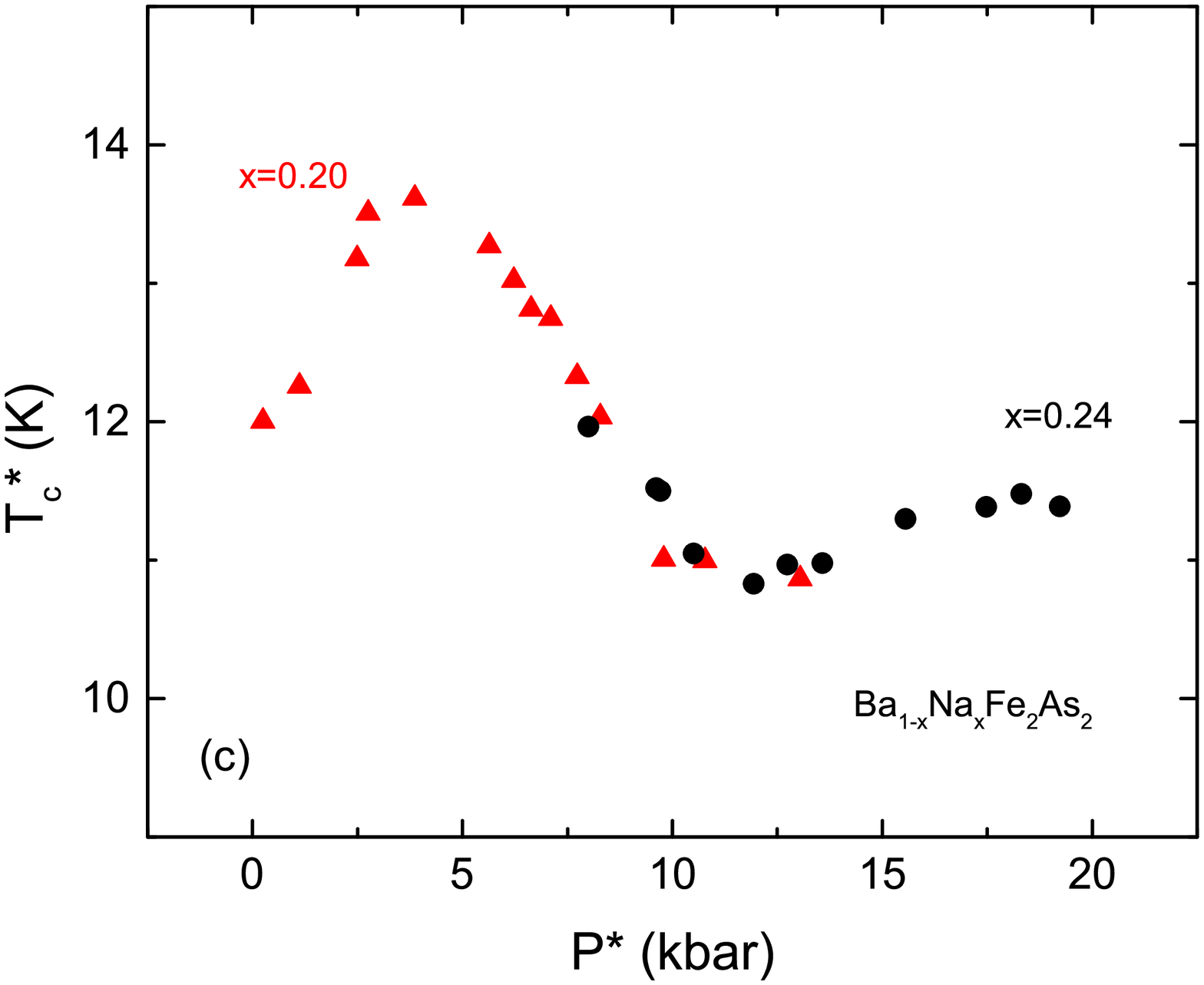}
\end{center}
\caption{(Color online) Pressure dependence of the superconducting transition temperature for (a)  Ba$_{0.8}$Na$_{0.2}$Fe$_2$As$_2$ and (b)Ba$_{0.76}$Na$_{0.24}$Fe$_2$As$_2$. Different symbols on panel (a) correspond to two different pressure runs. Numbers near the symbols on panels (a) and (b) correspond to the order in which the pressure was changed. Panel (c) shows both data sets with the one for $x = 0.24$ shifted by $+ 8$ kbar along the $X$ - axis and by $- 4$ K along the $Y$ - axis.}  \label{F5}
\end{figure}

On one hand, the unusual, non-monotonic behavior of the superconducting transition temperature under pressure for the  Ba$_{1-x}$Na$_x$Fe$_2$As$_2$  samples with $x = 0.2$ and $x = 0.24$ (Fig. \ref{F5})  could be considered to be consistent   with what one would expect for a Lifshitz transition. \cite{mak65a} At this moment the data on the Fermi surface of  Ba$_{1-x}$Na$_x$Fe$_2$As$_2$ \cite{asw12a} are regarded as very similar to those for  Ba$_{1-x}$K$_x$Fe$_2$As$_2$ \cite{evt09a,zab08a}, and no change of the Fermi surface topology, from the ARPES measurements, has been reported between the parent, BaFe$_2$As$_2$, compound and  Ba$_{1-x}$Na$_x$Fe$_2$As$_2$ with $x$ values up to 0.4, so the Lifshitz transition hypothesis seems unlikely.

On the other hand, an important feature, unique to the $x - T$ ambient pressure phase diagram of the  Ba$_{1-x}$Na$_x$Fe$_2$As$_2$ series, is the existence of a distinct, narrow antiferromagnetic and tetragonal, $C4$, phase. \cite{avc13a}  Both of the samples with non-monotonic pressure dependences of $T_c$ are located, at ambient pressure, close to the phase boundaries of this $C4$ phase. It is thus possible that the observed anomalies in $T_c(P)$ behavior for the  samples with $x = 0.2$ and $x = 0.24$  are associated with the crossing of these phase boundaries under pressure. If this supposition is correct, the effect of the $C4$ phase on superconductivity is different from that of the antiferromagnetic orthorhombic phase that is omnipresent in Fe-As materials, since crossing of the  antiferromagnetic orthorhombic phase line under pressure either has no effect on $T_c$ or this effect has previously been missed.

In summary, it appears that non-monotonic behavior of the $c$ crystallographic lattice parameter and the narrow antiferromagnetic tetragonal $C4$ phase both affect the pressure dependencies of the superconducting transition temperature in the Ba$_{1-x}$Na$_x$Fe$_2$As$_2$ series. Synthesis of homogeneous single crystals with finely controlled Na concentration  around 20-25\%  and further comprehensive measurements of superconducting and magnetic properties (and their interplay)  in this range of concentrations  would be desirable to understand the distinct properties of the  Ba$_{1-x}$Na$_x$Fe$_2$As$_2$ series.

\begin{acknowledgments}

We would like to thank Xiao Lin for assistance in the samples' handling, Adam Kaminski for discussion of published ARPES data, and S. Wurmehl and V. Grinenko for sharing their data from Ref. \onlinecite{gri13a}. Work at the Ames Laboratory was supported by the U.S. Department of Energy, Office of Science, Basic Energy Sciences, Materials Sciences and Engineering Division. The Ames Laboratory is operated for the U.S. Department of Energy by Iowa State University under contract No. DE-AC02-07CH11358. Work at  Argonne National Laboratory was supported by the U.S. Department of Energy, Office of Science, Materials Sciences and Engineering Division. 

\end{acknowledgments}

\clearpage

\clearpage

\clearpage

\clearpage

\clearpage

\clearpage

\clearpage

\end{document}